\documentclass[ reprint,
notitlepage, amsmath,amssymb,
 aps,
pre, 
]{revtex4-2}

\renewcommand{\exp}[1]{\mathrm{e}^{#1}}
\usepackage{mathtools}
\usepackage{graphicx,xcolor}
\usepackage{dcolumn}
\usepackage{bm}
\usepackage{hyperref}

\usepackage{nicefrac}

\let\vec\bm
\DeclareMathOperator{\grad}{\vec{\nabla}\!}
\newcommand{\ii}{\mathrm{i}}

\let\epsilon\varepsilon
\let\theta\vartheta
\newcommand\lldots{\makebox[1em][c]{.\hfil.\hfil.}}

\begin{document}

\title{Reduced density fluctuations via anti-aligning in active matter}

\author{Horst-Holger Boltz}
\email[]{horst-holger.boltz@uni-greifswald.de}

\author{Thomas Ihle}

\affiliation{Institute for Physics, Greifswald University, 
17489 Greifswald, Germany}

\date{\today}

\begin{abstract}
   We highlight the importance of long-range correlations in active matter systems of self-propelling particles even in the absence of global order or steric interactions by demonstrating that long-range density fluctuations are reduced. We show this analytically for a one-dimensional lattice process employing a Poisson representation. Within this framework, we are able to derive the fluctuating hydrodynamics for the Poisson fields. The emergent imaginary noise indicates the non-Poissonian nature of the number fluctuations and manifests in a non-trivial structure factor $S(k)$ which we are computing analytically. Numerically, we corroborate the relevance of these findings for off-lattice Vicsek-type models with anti-aligning interactions for which we observe apparent non-universal hyperuniformity which we suggest to interpret as a reduction with integer power-law to a finite value.  
\end{abstract}

\maketitle   

\section{Introduction} 

Active Matter physics, particulate systems driven out of equilibrium on the scale of particle size~\cite{marchetti2010,ramaswamy2010,chate2020}, is an intriguing field for exploring collective phenomena inaccessible in equilibrium physics; the seminal example being long-ranged orientational order~\cite{vicsek1995} in two dimensions. More generally, activity, especially intrinsic propulsion of particles, is a pathway to establish long-ranged correlations even in dilute systems with short-ranged interactions~\cite{hanke2013,chou2015,szamel2021,ihle2023short,ihle2023,kuersten2025,mihatsch2025}. This holds also in the absence of global spatial, orientational or other order. Taking this to its extreme, correlations between particles can still be relevant for the understanding of a particulate system even if the non-equilibrium steady-state (NESS) is characterized by a uniform single-particle distribution function. These correlations quantify the deviation of the statistics of positional configurations from that of realizations of a simple uniform Poisson point process. A special class of disordered point patterns are hyperuniform configurations~\cite{torquato2018}, which show anomalous scaling behavior in the number fluctuations of large subsystems\footnote{Such fluctuations do not follow a volume law, whereas a Poisson point process would show $\Delta N {\sim} N {\sim} r^d$.}, i.e. $\Delta N \sim r^\beta$ for $r\to \infty$ with $d{-}1{<}\beta{<}d$. A complementary definition of hyperuniformity is that the static structure factor $S(k)=\langle \delta\tilde\rho(k)\, \delta\tilde\rho({-}k)\rangle$ approaches zero in the long-wavelength limit, $S(k{\to}0)\to 0$. Recently, it has been shown that a broader class of field-theories of active particle systems generically lead to hyperuniformity.~\cite{zheng2023} On the micro- or mesoscopic level, active or generally non-equilibrium processes that lead to hyperuniform configurations include \cite{lei2024} convection\cite{weijs2017,gilpin2018, lejeune2023}, chiral self-propelled-particles~\cite{lei2019b,huang2021,zhang2022,oppenheimer2022,kuroda2023,chen2024}, Manna-processes with center-of-mass conservation~\cite{hexner2017,bertrand2019,mukherjee2024} and adsorbing states~\cite{hexner2015,wang2018,lei2019,ma2019,ma2023}. For the latter two, hyperuniformity can be understood as a consequence of how density fluctuations are resolved in the problem non-locally.

Intuitively, a similar reasoning applies to systems of self-propelled particles at low densities that feature short-ranged anti-aligning interactions~\cite{ihle2023,ihle2023short,menzel2012,chatterjee2023,escaff2024,escaff2025,escaff2025b,musacchio2025,lardet2025,kuersten2025}. Anti-aligning orientational interactions are less prominent than their aligning counterparts, but some experimental work (albeit in combination with other  interactions) exists~\cite{nourhani2021,das2024}.  For low densities, a density fluctuation will lead to particles interacting and, typically, to opposed orientations. Thus, center-of-mass motion, while not completely inhibited by conservation, is suppressed, as are successive interactions of the same particles.  We focus on deterministic models as this effect would be counteracted by both orientational and spatial noise.
 
In this article, we follow the intuition and find an active matter mechanism that reduces long-range density fluctuations. We perform a Poisson representation analysis of a one-dimensional model active lattice gas~\cite{thompson2011, manacorda2017, bertrand2019, benvegnen2022} with {\em anti-aligning} interactions such that the dynamics reduce local polarization in the directions of motion. We find that correlated noise in a collective description does lead to a significant reduction of large-scale density fluctuations and extend this observation to the more commonly studied case of two dimensions by means of agent-based simulations. In particular, we study numerically an anti-aligning continuous-time variant~\cite{peruani2008} of the seminal Vicsek model~\cite{vicsek1995}, an important toy model for the {dry active aligning dilute matter paradigm}~\cite{chate2020}.  From these simulations, we report numerical evidence of apparent anomalous hyperuniformity.

\section{pedagogical one-dimensional model} 
\begin{figure}
\includegraphics{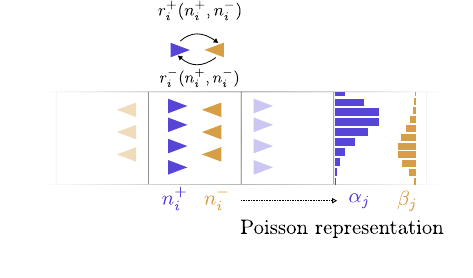}
\caption{Graphical representation of the one-dimensional model. The agent-based model consists of particles on a lattice that move respective to their internal orientation to an adjacent lattice site (left part, possible motion indicated by shaded particles). In the analytical treatment, we move from the integer occupation numbers $n_i^\pm$ to complex-valued Poisson fields (right part), $\alpha_j$ and $\beta_j$, which control the local number statistics. The Poisson fields themselves are random quantities from a distribution $f$. Any observable moment is to be inferred by averaging over all realisations of the Poisson fields and, thus, real.} \label{fig:psketch}
\end{figure}

We consider a one-dimensional lattice populated by two species of particles that are moving in continuous time to the right ($+$) or left ($-$) subject to a process that changes the species of particles. We denote the number of particles of species $\pm$ on lattice site $i$ by $n_i^\pm$. The dynamics is then given by two processes, streaming and conversion: Particles of species $\pm$ move from lattice site $i$ to $i\pm1$ with constant rate $\lambda$ and convert to species $\mp$ with rates $r^\pm (n_i^+, n^-_i)$. Thus, we assume that any changes in particle species are only due to local effects. A graphical representation is given in Fig.~\ref{fig:psketch}.  This model is close in spirit to previously discussed models, e.g., the one in ref.~\citenum{bertrand2019}. However, we discuss explicitly active particles whose microscopic mobility is unaffected by the density. Instead we will see that a similar phenomenology can be achieved effectively by coupling the conversion to the local orientational statistics. {Similarly, there is an inevitable overlap with the active Ising model~\cite{solon2013,*solon2015}, but here transport is strictly convective, allowing for only one direction of motion per species, and athermal as the conversion rates that we are using are not of Arrhenius type.}

We can formulate a master equation for the probability distribution $P(\{n_i^+,n_i^-\})$ to find a specific set of occupancy numbers $\{n_i^+,n_i^-\}$
\begin{equation}
    \begin{split}
   & \frac{\partial P(n_1^+,n_1^-,n_2^+,n_2^-,\lldots,\lldots)}{\partial t}= \frac{\partial P(\{n_i^+,n_i^-\})}{\partial t}\\
    &= - (S^\lambda_\text{out} + C^r_\text{out}) + S^\lambda_\text{in}+C^r_\text{in}
    \end{split}
\end{equation}
with the probability outfluxes (using additional superscripts to highlight the parameter dependence) due to streaming $S^\lambda_\text{out}$ and conversion $C^r_\text{out}$, respectively,
\begin{subequations}\label{eq:operators1}
\begin{align}
    S^\lambda_\text{out}&=\sum_i \left[ \lambda n_i^+ {+}\lambda n_i^-  \right] P \\ 
    C^r_\text{out}&=\sum_i \left[  r^+ n^+_i\! +\! r^- n^-_i \right] P 
\end{align}%
\end{subequations}%
and the analogous influxes (for streaming we write first the influx from the left and then from right, for conversion first from plus to minus)
\begin{subequations} \label{eq:operators2}
\begin{align}
    S^\lambda_\text{in}&=\sum_i \left[ \lambda (n_{i-1}^++1) P(\lldots,n_{i-1}^+{+}1,n_{i-1}^-,n_{i}^+\!{-}\!1,\lldots) \right.\notag\\
    &+\left.\lambda ( n_{i+1}^-{+}1) P(\lldots, n_{i}^-\!-\!1,n_{i+1}^+, n_{i+1}^-{+}1,\lldots) \right] \\
    C^r_\text{in}&=\!\sum_i\! \left[   r^+(n_i^+{+}1,n^-_i\!{-}\!1) (n_i^+{+}1) P(\lldots, n_i^+{+}1,n^-_i\!{-}\!1,\lldots)\right.\notag\\
    &\!\!\!\!\!\!\!\!\!\!\!\!{+}\left. r^-(n_i^+\!{-}\!1,n_i^{-}{+}1) (n_i^-{+}1) P(\lldots, n_i^+\!{-}\!1,n^{-}_i{+}1,\lldots) \right]  \text{,}
\end{align}
\end{subequations}

where omitted arguments on the right-hand-side are identical to the ones on the left in the distribution $P$ as well as for $r^\pm(n_i^+,n_i^-)$.

Solving the Master equation as it is, is not a fruitful endeavor, since the right-hand side is nonlinear for interacting systems, i.e non-trivial conversion rates. We use a representation in terms of local Poisson processes~\cite{gardiner1977,*chaturvedi1978,*gardiner1985} with local rates $\alpha_i$ and $\beta_i$, see also Fig.~\ref{fig:psketch}, for the two species. The idea of this longly established methodology is to move from a discrete stochastic process in positive integers (here the occupation numbers $n_i^\pm$) to a continuous process and similarly move from Master equations to a Fokker-Planck-equation. The new continuous variables (here the rates $\alpha_i$ and $\beta_i$) act as the parameters of a Poisson process that then generates the occupation numbers $n_i^\pm$. Statistical moments are therefore acquired by averaging over the realisations of the Poisson process at fixed fields and their distribution. As a direct consequence of the mismatch in generality between a general discrete process and local Poisson processes, a higher level of abstraction is needed: for one, the local rates are complex numbers and, for two, the distribution is really a pseudodistribution as it also can be locally negative. Neither of these extensions are unphysical as the Poisson fields are auxialiary mathematical objects that are not directly accessible. Accessible are moments of the physical process, which are positive and real. Imaginary contributions to the Poisson fields can be seen as a deviation from Poissonian statistics, which manifestly couple the mean and the variance of a random variable. To understand this intuitively, one can compare a Poissonian variable $n$ with rate $\lambda$ to a variable $m$ with rate $\lambda+\ii \xi$ wherein $\xi$ is a (for the Poisson process quenched) random variable with $\overline{\xi}=0, \overline{\xi^2}=c$. Naive adapation of the textbook results for a Poisson process leads to $\langle n\rangle=\langle n^2\rangle_c = \langle m \rangle = \lambda$, but $\langle m^2\rangle_c =\lambda^2-c$. Imaginary noise as an unintuitive consequence of the Poisson representation is more thoroughly discussed in the literature~\cite{wiese2016,correales2019} and will be remarked upon later where relevant.  The strength of working in Poisson representation is a direct, non-asymptotic access to the fluctuations (which will result in a noise term in the final Langevin equation) and their correlations which are crucial to determine the structure factor. This is in contrast to some path integral approaches that derive uncorrelated Gaussian noise asymptotically.~\cite{lefevre2007,thompson2011}

We denote the distribution of the Poisson rates $\alpha_i,\beta_i$ by $f=f(\{\alpha_i,\beta_i\})$ and it relates to the distribution $P=P(\{n_i^\pm\})$ of the occupation numbers via
\begin{align}
    P&= \!\int\!\!\!\ldots\!\!\!\int\! \prod_i \left(\! \mathrm{d}\alpha_i\, \mathrm{d}\beta_i\,  \frac{\exp{{-}\alpha_i}\alpha_i^{n_i^+}}{(n_i^+)!} \frac{\exp{{-}\beta_i}\beta_i^{n_i^-}}{(n_i^-)!}\! \right)f \text{.} \label{eq:poisson}
\end{align}
An equation of motion for $f$ is directly available by inserting this representation for $P$ into the master equation. For the occurring integrals , we assume boundary conditions such that any boundary term in integration by parts vanish.  We are interested in anti-aligning interactions, therefore we choose on the simplest level
\begin{align}
r^{\pm}(n^+,n^-) = r_0 + r_1 \cdot (n^{\pm} - 1)\label{eq:rate}
\end{align}

Because we are interested in the conversion due to present magnetization, the important parameter is $r_1$. We keep the ``spontaneous conversion rate'' $r_0$ throughout for generality, but are interested in $r_1$ dominating over $r_0$. In the specific version of eq.~\eqref{eq:rate}, we are explicitly discarding self-interactions. { Other conversion models are imaginable. However, terms higher than $\mathcal{O}(n^+,n^-)$ in $r^{\pm}$ would lead to higher than second derivatives in the equation of motion for the pseudo-probability-density-function of $f$ which, therefore, would not be of Fokker-Planck type and could not be solved (as we will do below) by directly solving a corresponding Langevin equation with exact expressions for the relevant noise (which is an important quality of Poisson representations~\cite{kourbane2018}). The conversion rate of eq.~\eqref{eq:rate} is special in this regard and used throughout.}

\subsection{Details of Poisson representation}
Using the definition of the Poisson representation eq.~\eqref{eq:poisson}, we get the following rules for replacing terms by means of integration by parts and identification of terms. In the following, we are omitting all the arguments of $f$ and $P$ whenever they coincide with the ``default'' values $\{\alpha_i,\beta_i\}$ or $\{n_i^+,n_i^-\}$, respectively. Similarly, we only write out the relevant parts of the Poisson representation. Finally, we reiterate that $f$ can be constrained to functions such that all boundary terms in integration by parts (marked by $p.I.$) vanish. For streaming (and ``spontaneous'' conversion due to $r_0$) contributions, the prototypical linear terms are 
\begin{align}
\begin{split}    n_i^+ P =&  \int\!\! \ldots\! \!\int\!\mathrm{d}\alpha_i\,\frac{\exp{{-}\alpha_i}\alpha_i^{n_i^+}}{(n_i^+)!} n_i^+ f\\=&\int\!\! \ldots\! \!\int\!\mathrm{d}\alpha_i\,\frac{\alpha_i^{n_i^+-1}}{(n_i^+-1)!} \exp{{-}\alpha_i} \alpha_i f\\
    \stackrel{p.I.}{=}& \int\!\! \ldots\! \!\int\!\mathrm{d}\alpha_i\,\frac{\alpha_i^{n_i^+}}{(n_i^+)!} (-\partial_{\alpha_i}) \exp{{-}\alpha_i} \alpha_i f \\=& \int\!\! \ldots\! \!\int\!\mathrm{d}\alpha_i\,\frac{\alpha_i^{n_i^+}}{(n_i^+)!} \exp{{-}\alpha_i} (1-\partial_{\alpha_i}) \alpha_i f \end{split}\end{align}
   for streaming to the right and
\begin{align}
    \begin{split}&(n_{i-1}^++1) P(\ldots,n_{i-1}^+{+}1,\ldots,n_i^+{-}1,\ldots)\\&=\int\!\! \ldots\! \!\iint\!\mathrm{d}\alpha_{i-1}\,\mathrm{d}\alpha_i\,\frac{\exp{{-}\alpha_{i-1}{-}\alpha_i}\alpha_{i-1}^{n_{i-1}^++1}\alpha_i^{n_i^+-1}}{(n_{i-1}^+)! (n_i^+-1)!}  f\\
    &=\int\!\! \ldots\! \!\int\,\mathrm{d}\alpha_i\,\frac{\exp{{-}\alpha_i}\alpha_i^{n_i^+-1}}{(n_i^+-1)!} \alpha_{i-1}^+ f \\&\stackrel{p.I.}{=}\int\!\! \ldots\! \!\int\,\mathrm{d}\alpha_i\,\frac{\exp{{-}\alpha_i}\alpha_i^{n_i^+}}{(n_i^+)!} (1-\partial_{\alpha_i})\alpha_{i-1}^+ f \text{ } \end{split}
\end{align}
for streaming from the left. By symmetry, we can extend this to the following replacement rules that will eventually allow to map the master equation to a Fokker-Planck-equation. We find
\begin{align}
    \sum_i n_i^+ P \to& \sum_i (\alpha_i f - \partial_{\alpha_i} \alpha_i f)\\
     \sum_i n_i^- P \to& \sum_i (\beta_i f - \partial_{\beta_i} \beta_i f)\end{align}
     and \begin{align}
        \begin{split}
   & \sum_i (n_{i-1}^++1) P(\ldots n_{i-1}^++1, \ldots, n^+_i-1,\ldots) \\ &\qquad \to \sum_i (\alpha_{i-1} - \partial_{\alpha_i} \alpha_{i-1}) f\end{split}\\
    \begin{split}&\sum_i (n_{i+1}^-+1) P(\ldots n_{i}^--1, n^-_{i+1}+1,\ldots) \\ &\qquad \to
    \sum_i (\beta_{i+1} - \partial_{\beta_i}\beta_{i+1}) f\text{.}\end{split}
\end{align}
For the conversion, we consider the following non-linear terms. Conversion of plus to minus is represented by
\begin{align}
    \begin{split}
    &n_i^+ (n_i^+-1) P \\=&  \int\!\! \ldots\! \!\int\!\mathrm{d}\alpha_i\,\frac{\exp{{-}\alpha_i}\alpha_i^{n_i^+}}{(n_i^+)!} n_i^+ (n_i^+-1)  f \\ \stackrel{p.I.}{=}& \int\!\! \ldots\! \!\int\!\mathrm{d}\alpha_i\,\frac{\alpha_i^{n_i^+}}{(n_i^+)!} \partial_{\alpha_i}^2 \exp{{-}\alpha_i} \alpha_i^2 f \\=& \int\!\! \ldots\! \!\int\!\mathrm{d}\alpha_i\,\frac{\alpha_i^{n_i^+}}{(n_i^+)!} \exp{{-}\alpha_i} (1-\partial_{\alpha_i})^2 \alpha_i^2 f \end{split} \end{align} and conversion from minus to plus by \begin{align}
    \begin{split}&(n_{i}^++1) n_i^+ P(\ldots,n_{i}^+{+}1,n_i^-{-}1,\ldots)\\ =&\int\ldots\iint\!\mathrm{d}\alpha_i\,\mathrm{d}\beta_i\, \frac{\exp{{-}\alpha_i}\alpha_i^{n_i^++1}}{(n_i^+-1)!} \frac{\exp{{-}\beta_i}\beta_i^{n_i^--1}}{(n_i^--1)!} f \\ \stackrel{p.I.}{=}& \int\!\! \ldots\! \!\int \!\mathrm{d}\alpha_i\,  (1-\partial_{\beta_i})\frac{\exp{{-}\alpha_i}\alpha_i^{n_i^+-1}}{(n_i^+-1)!} \alpha_i^2 f\\\stackrel{p.I.}{=}&\int\!\! \ldots\! \!\int \!\mathrm{d}\alpha_i\, \frac{\exp{{-}\alpha_i}\alpha_i^{n_i^+}}{(n_i^+)!} (-1+\partial_{\alpha_i})(-1+\partial_{\beta_i}) \alpha_i^2f\end{split}
\end{align}
and we, therefore, find the following replacement rules
\begin{align}
    &\sum_i (n_i^+-1) n_i^+ P \to \sum_i (1-\partial_{\alpha_i})^2 \alpha_i^2 f\\
    &\sum_i (n_i^--1) n_i^- P \to \sum_i (1-\partial_{\beta_i})^2 \beta_i^2 f\\
    \begin{split}&\sum_i (n_i^++1) n_i^+ P(\ldots, n_i^++1, n_i^--1,\ldots) \\&\to \sum_i (-1+\partial_{\alpha_i})(-1+\partial_{\beta_i}) \alpha_i^2f \end{split}\\    
    \begin{split}&\sum_i (n_i^-+1) n_i^- P(\ldots, n_i^+-1, n_i^-+1,\ldots) \\&\to \sum_i (-1+\partial_{\alpha_i})(-1+\partial_{\beta_i}) \beta_i^2f \text{.}\end{split}
\end{align}
Some of these terms  cancel each other out when considering the relevant combinations from eqs.~\eqref{eq:operators1} and \eqref{eq:operators2}. Gathering terms, we find the Fokker-Planck-equation given in the following section.

\subsection{Fokker-Planck equation}
The Fokker-Planck equation for the distribution $f$ of the local Poisson parameters is then
\begin{subequations} \label{eq:fp}
\begin{align}
\partial_t f &= F_\lambda + F_r + J
\end{align}
with contributions of deterministic origin $F_{\lambda,r}$ and a noise term $J$ (in the implied Langevin equation)
\begin{align}
    \begin{split}
       F_\lambda &= \lambda \sum_i [(\alpha_{i-1} - \alpha_i)  + \partial_{\alpha_i} (\alpha_i-\alpha_{i-1})]f- \\&\lambda \sum_i [ (\beta_{i}-\beta_{i+1}) + \partial_{\beta_i} (\beta_{i+1}-\beta_{i}) ]f\end{split}\\
       \begin{split}
       F_r &=r_0 \sum_i (\partial_{\alpha_i}-\partial_{\beta_i})(\alpha_i-\beta_i) f\\&+r_1 \sum_i (\partial_{\alpha_i} -\partial_{\beta_i}) (\alpha+\beta)(\alpha-\beta) f\end{split}\\
       \begin{split}
       J&=-r_1 \sum_i [ \partial^2_{\alpha_i} \alpha_i^2 +  \partial^2_{\beta_i} \beta_i^2 \\
       &\qquad\qquad- \partial_{\alpha_i}\partial_{\beta_i} (\alpha_i^2+\beta_i^2) ]f \text{.} \end{split}
\end{align}    
\end{subequations}

The last line contains terms with second derivatives, i.e. noise in a corresponding Langevin equation. The deterministic equations of motion for $\alpha(x)$, $\beta(x)$ are
\begin{subequations}
\begin{align}
    \partial_t \alpha\rvert_\text{det} &= {-}v \grad \alpha - r_0 (\alpha {-} \beta) - r_1 (\alpha{+}\beta)(\alpha{-}\beta) \\
    \partial_t \beta\rvert_\text{det} &= v\grad \beta + r_0 (\alpha-\beta) + r_1 (\alpha+\beta)(\alpha-\beta) \text{.}
\end{align}
\end{subequations}

While we switched our notation from $\lambda$ to $v$ to be more in line with conventions, $\grad$ is used here as a short-hand for next-neighbor differences, no continuum limit has been performed. This is significant below.

\subsection{Noise}
To determine the noise terms the Fokker-Planck equation eq.~\eqref{eq:fp} into the standard form
\begin{align}
    \partial_t f=&- \sum_i \mu_i f + \sum_{i,j} \frac{\partial^2}{\partial x_i\partial x_j} \left[ D_{ij} f \right] \text{,}
\end{align}
we find the diffusivity matrix $D_{ij}$ to be
\begin{align}
    D_{ij}&=-r_1\begin{pmatrix}
        \alpha_i^2 & - (\alpha_i^2+\beta_i^2)/2\\
        -(\alpha_i^2+\beta_i^2)/2 & \beta_i^2
    \end{pmatrix} \\\begin{split}&=  -r_1 \left[\frac{\tilde\rho^2}{4}\begin{pmatrix} 1 & -1\\ -1 & 1\end{pmatrix} + \frac{\tilde\rho \tilde m}{2} \begin{pmatrix}1&0\\0&-1\end{pmatrix} \right. \\ &\qquad\qquad \left.+ \frac{\tilde m^2}{4} \begin{pmatrix}1&0\\0&1\end{pmatrix} \right] \text{,} \end{split}
\end{align}
wherein we have used $\tilde \rho=\alpha_i+\beta_i$, $\tilde m=\alpha_i-\beta_i$. Physically, we are interested in the limit $\tilde \rho \approx \rho_0 \gg m\approx 0$. To translate the Fokker-Planck equation into a Langevin equation we have to compute the matrix radical $B=\sqrt{2D}$. To zeroth order in $\tilde m$, we find
\begin{align}
B&=\frac{\ii \rho \sqrt{r_1}}{2} \begin{pmatrix}1&-1\\-1&1\end{pmatrix}
\end{align}
from which it becomes apparent that the two eigenvalues are $\lambda_\rho=0$ and $\lambda_m=\ii \sqrt{r_1}\rho$. Here, it is important to understand the significance of an imaginary noise strength.~\cite{wiese2016,correales2019} As we work in the Poisson representation, $\tilde \rho=\alpha+\beta$ is not directly the (scaled) density even in a continuum limit (in the sense of an empirical measure) on a strict semantic level, but the parameter of a Poisson process whose first moment is the density, which is why we mark the combinations of $\alpha$ and $\beta$ as $\tilde\rho$ and $\tilde m$, respectively. As such there would be physical density fluctuations, even if the dynamics of $\tilde\rho$ were deterministic, but it would be a Poisson process with mean and variance being identical. Arbitrary deviations from this Poissonian nature are only possible if the Poisson parameter is complex; whence the appearance of an imaginary noise. Specifically, an imaginary noise leads to a narrowing of the microscopic distributions, whereas real noise would imply a widening.~\cite{wiese2016} Importantly, this means that fluctuations of our Poisson parameter $\tilde\rho$ are not identical to the fluctuations of the number of particles, however we can relate them for the quantity of our interest, the structure factor.

\subsection{Langevin equation}
To summarize: Inspecting the diffusivity matrix in the relevant case $\tilde\rho=\alpha+\beta \gg \lvert \tilde m \rvert$ with $\tilde m = \alpha-\beta\approx 0$, we find that its two eigenvalues are $\lambda_{\tilde m}=\ii \sqrt{r_1}\tilde \rho/2$ and $\lambda_{\tilde\rho}=0$ where we used subscripts to indicate the action of the corresponding eigenvectors. {We stress that $\tilde m,\tilde \rho$ are the rates of Poissonian processes whose mean are the physical quantities $m, \rho$ and not the same quantity.} There is no noise in the dynamics of $\tilde\rho$: in the Poissonian representation, noise  originates solely from non-linear terms. Since the only non-linear terms in the dynamics are in the conversion, which is strictly local, and preserves the total local occupancy, this still holds. Consequently, we arrive at a set of coupled Langevin equations
\begin{subequations} \label{eq:fp2}
\begin{align}
    \partial_t \tilde \rho &= - v\grad \tilde m \\
    \partial_t \tilde m &= -v \grad \tilde \rho - r_0 \,\tilde m - r_1 \,\tilde\rho \, \tilde m + \ii \sqrt{r_1} \,\tilde\rho \,\xi \text{.}
\end{align}
\end{subequations}
Here, $\xi$ denotes a Gaussian uncorrelated noise, with $\langle \xi\rangle =0$ and $\langle \xi(t)\xi(t')\rangle = \delta(t-t')$.  We reiterate that the emergence of an imaginary noise in the dynamics of the Poisson fields is a well-known hallmark~\cite{wiese2016,correales2019} of non-Poissonian statistics {that break the equality of mean and variance present in real Poisson processes}. It does not imply any imaginary observables.

The relevance of our model is corroborated by observing that the deterministic version of this set of equations shares the same structure (with $r_0\equiv 0$) as closed hydrodynamics of anti-aligning particles~\cite{patelli2021,ihle2023,pinto} when rephrased in a one-dimensional settings. While general Toner-Tu like approaches~\cite{toner1998,*toner2005} cover these dynamics as well, they do not make predictions about $r_0$ and $r_1$, respectively.

We want to consider the dynamics of small perturbations around a uniform, non-polarized state. We decompose $\tilde\rho=\rho_0 + \delta\!\tilde\rho$ and treat both $\delta\!\tilde\rho$ and $\tilde m$ as small. Additionally {and based on physical intuition}, we reduce the multiplicative noise to an additive one, i.e. we ignore a term $\propto \delta\!\tilde\rho\, \xi$ which is not directly second order in the sense of our approximation, but should be negligible provided $\delta\!\tilde\rho$ is sufficiently small, as required by consistency. This approach is rather common in analytical studies due to intricacies of treating multiplicative noise. A prominent class of problems that are routinely approximated this way is related to applications of the Dean equation.~\cite{dean1996,illien2025} With these preliminaries, we rewrite eq.~\eqref{eq:fp2} as a single, closed second-order equation
\begin{align}
\partial_t^2 \delta\!\tilde\rho = v^2 \Delta \delta\!\tilde\rho - (r_0 + r_1 \rho_0) \partial_t \delta\!\tilde\rho - \ii \sqrt{r_1} \rho_0 v \grad \xi \text{.}
\end{align}

Inspecting the deterministic dynamics in Fourier space (transforming from position $x$ to wave-number $k$, {indicated by a subscript of k}), we find a harmonic oscillator. In the relevant limit, $k\to 0$, it is overdamped: both dynamical eigenvalues are negative. The long-time behavior is  dominated by the larger one, which is
$
    \mu = - \frac{v^2 k^2}{r_0 + r_1 \rho_0}
$. The diffusive scaling, $\sim \exp{- D k^2 t}$, results in a diffusivity $D=v^2/(r_0+r_1 \rho_0)$. As the deterministic dynamics are overdamped on large scales, we can neglect the inertial term. This way, we have reduced everything to one first-order Langevin or noisy diffusion equation
\begin{align}
\partial_t \delta\!\tilde\rho_k &= -D k^2 \delta\!\tilde\rho_k + \frac{\ii v \sqrt{r_1}  \rho_0 \eta_k}{r_0 + r_1 \rho_0} = -D k^2 \delta\!\tilde\rho_k + \ii c \eta_k  \label{eq:wnoise}
\end{align}

{Here, we introduced the shorthand $c= \frac{v \sqrt{r_1}}{r_0 + r_1 \rho_0} \rho_0$, which quantifies the strength of the noise.}
To understand the statistics of the effective noise in Fourier-space $\eta_k$ introduced here, it is important to reiterate that $\grad \xi$ should really be read as $\xi_{i+1} - \xi_{i}$. To account for this, we introduce a drift-free correlated noise~\cite{bertrand2019} $\eta_i=\eta(i)$ with \begin{align}
\langle \eta_i(t) \eta_j (t')\rangle &= \delta(t-t') \cdot \begin{cases} 2 & i=j\\ -1 & i=j \pm 1 \\ 0 & \text{else.} \end{cases}
\end{align}
 
In Fourier space, the relevant noise correlations are
\begin{align}
\langle \eta_k (t) \eta_{k'}(t') \rangle &= \delta(k+k') \delta(t-t') \, 2(1-\cos(k)) \text{.}
\end{align}
With this correlated noise, we can solve eq.~\eqref{eq:wnoise}~\cite{hexner2017,bertrand2019} and find the structure factor, $N S(k)=\langle \rho_k\rho_{-k}\rangle=\rho_0+\langle \delta\!\tilde\rho_k\delta\!\tilde\rho_{-k}\rangle$, at large times
\begin{align}
    N S(k)&= \rho_0 - \frac{c^2}{D k^2}(1-\cos(k)) \\
    &= \rho_0 - \frac{c^2}{2D} (1-\frac{k^2}{12} +\ldots)\text{.} \label{eq:pred}
\end{align}
Here, we used the important result that the structure factor, which is a descriptor of the physical densities $\rho$, can indeed be computed from the fluctuations of the associated Poisson field.~\cite{bertrand2019} Comparing $S_0=\lim_{k\to0^+} N S(k) = \rho_0 - c^2/(2D)$  with the Poissonian value $S\rvert_{r\equiv0} = \rho_0$, we find
\begin{align}
 \frac{S_0}{\rho_0}&=1-\frac{c^2}{2 D \rho_0} = 1-\frac{r_1 \rho_0}{2(r_0+r_1\rho_0)} \text{.}
\end{align}

In the relevant limit of interactions being dominant $r_1\rho_0 \gg r_0$, this approaches $S_0\rho^{-1}_0 = 1/2$. A suitable mechanism to incorporate a spontaneous conversion $r_0$ into typical microscopic models would be the introduction of orientational noise into the model corroborating the earlier statement that the dynamically built-up correlations will be more present in deterministic models.  We note that lower values, including the hyperuniform case of $S_0\to 0$, might be possible with different models particularly if center-of-mass~\cite{hexner2017,bertrand2019,mukherjee2024} or polarization are strictly conserved. This idea has also been expressed in Ref.~\citenum{maire2025} during the preparation of this manuscript. Here, we are more interested in the generic effect of anti-aligning interaction in an actively moving matter on the large scale statistics.

\begin{figure}
    \includegraphics[scale=0.5]{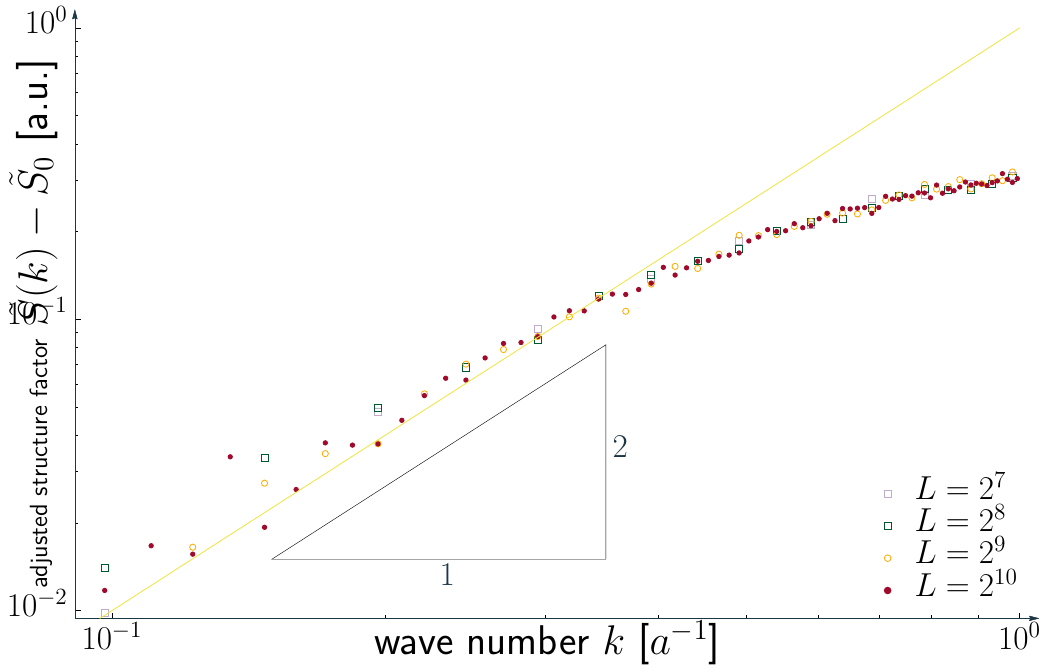} 
    \caption{Structure factor {as a function of the wave number (in units of inverse lattice constants)} from a numerical realization of the one-dimensional active lattice gas model. Here, we operated directly in the limit $r_1 \gg r_0$ as the conversion rate is directly proportional to $m$. We provide guides to the eye to show the dominant power-law behaviors for small $k$. {Shown is the measured structure factor rescaled to account for the difference between binomial and Poissonian statistics in a finite system and reduced by predicted offset $S_0$.}}
    \label{fig:sf1d}
\end{figure} 

We present data from a numerical realization of the proposed stochastic process in fig.~\ref{fig:sf1d}. We find our data to be in qualitative agreement with the prediction of eq.~\eqref{eq:pred} for sufficiently large system sizes, indicating a reduction of long-range density fluctuations in this system. The agreement is qualitative in the sense that we use the gestalt of the result but currently are not able to predict the offset and the prefactor to the $k^2$-term.

\begin{figure}
    \includegraphics[scale=0.5]{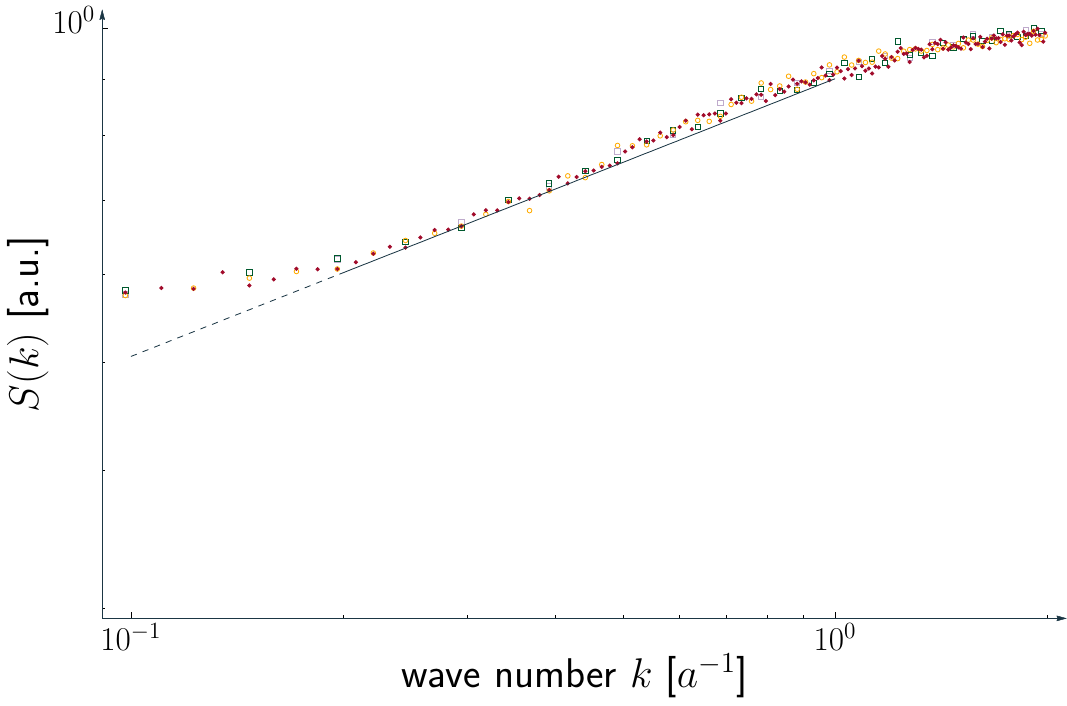}
    \caption{Raw data of the structure factor (symbols and colors being identical to the ones used {in Fig.~\ref{fig:sf1d}}) with a direct power-law line as a fictitious inference of anomalous hyperuniformity. In this case, we would find $k^\alpha$ with $\alpha\approx 0.25$. For the simple, one-dimensional model, this works only on a small-range of $k$-values spanning roughly half a decade.}
    \label{fig:sf1d_art}
\end{figure}

We consider a one-dimensional lattice of size $L$ with periodic boundary conditions. Each lattice site has a occupancy for two species: left- and right-moving particles. Initially the lattice is filled such that the average total number (adding both species) is $\rho_0=16$. {We integrate the dynamics by means of Gillespie's algorithm~\cite{gillespie1976,*gillespie1977}}: At each lattice site there are three rates to evaluate: a particle will move at total rate $R_i^\lambda= \lambda (n_i^++n_i^-)$, a conversion will happen with total rates $R_i^\pm = r_1 (n_i^\pm -1 )n_i^\pm$. From these the time to the next reaction is computed as a random number drawn from an exponential distribution, $\delta t \sim \text{EXP}(\tau=(\sum_i \sum_j R_i^j)^{-1})$. Finally, the specific reaction is chosen randomly using the corresponding rates as relative weights.

The results of this procedure are analyzed by evaluating the one-dimensional structure factor after a time of at least $t=128$ has passed (in units such that $\lambda=1, r_1=1/100$) and averaging over $10^4$ samples. The quantity shown in {Fig.  \ref{fig:sf1d}} is a rescaled structure factor (corresponding to densities scaled by $(\rho_0^2+\rho_0)/(\rho_0^2)$) with the expected $y$-intercept of $0.5$ subtracted.

In Fig.~\ref{fig:sf1d_art}, we demonstrate directly the {main argument} for the data in the one-dimensional model. Ignoring or being ignorant of the offset at small $k$ could possibly lead to the inference of an apparent power-law, in this case $k^{\approx 0.25}$.

Anti-aligning ``colliding'' interactions in one dimension are fundamentally peculiar as the only way of anti-aligning colliding particles via local interactions, flipping a pair of opposedly aligned particles,  leads to identical configurations for indistinguishable particles. As such, it is not possible to implement a local interaction rule that would conserve $m$. This is crucially different in higher dimensions, where the non-conservedness of the momentum that is genuine to active particles would also allow for an additional rotation of the pair. Unfortunately, it is unclear how to extend this analytical description to higher dimensions, as commensurability to the spatial lattice enforces a discrete symmetry in the orientational degree of freedom which will affect the collective behavior~\cite{benvegnen2023}.  Leaving these avenues to potential future work, we proceed by exploring the more relevant off-lattice two-dimensional case numerically.

\section{Two-dimensional Vicsek-like model}
 We consider a continuous-time variant~\cite{peruani2008} of the paradigmatic  Vicsek model~\cite{vicsek1995,ginelli2016} of self-propelled particles with anti-alignment interactions. Describing a particle's orientation by the angle $\theta_i$ ($i=1,\ldots,N$) to a fixed reference axis, the positional dynamics are given by streaming into this direction whereas the alignment interactions are done via short-ranged Kuramoto-like couplings
\begin{subequations}\label{eq:model}
\begin{align}
\dot{\vec r_i} &= v_0 \vec n(\theta_i)\\
\dot{\theta_i} &= -\Gamma \sum\nolimits_{j\in \Omega_i} \sin{(\theta_j-\theta_i)} \text{.}
\end{align}%
\end{subequations}%
Here, $\vec n(\theta)=(\cos\theta,\sin\theta)^\mathrm{T}$ is the unit vector in the direction corresponding to $\theta$, and $\Omega_i$ is the set of indices of particles within a radius $R$ around $\vec r_i$, i.e. $\Omega_i=\{j\,|\,j\neq i \,\&\, \lvert \vec r_i -\vec r_j\rvert \leq R\}$. This model has originally been used to study flocking~\cite{peruani2008,kuersten2021} with aligning {($\Gamma <0$)} interactions. 

While the methodology based on the assumption of {\em one-sided} molecular chaos~\cite{kreuzer1981} used for a recent quantitatively successful analytical description of the anti-aligning case~\cite{ihle2023short,ihle2023} does not neglect two-particle correlations, they are only considered during and on the scale of interactions and, thus, there is no direct way to infer a prediction for the structure factor from this asymptotically exact solution. Tracking long-ranged correlations analytically remains challenging, so we resort to straightforward agent-based simulation of eq.~\eqref{eq:model} which is feasible with existing~\cite{kuersten2023} code.
   
\begin{figure}
    \includegraphics[scale=0.5]{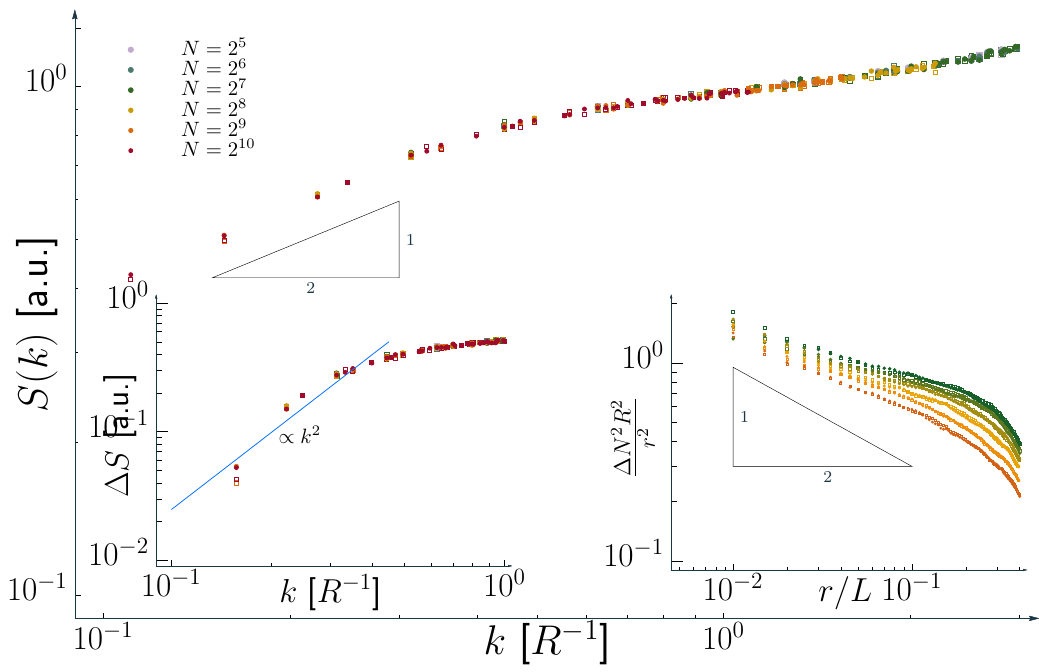}
    \caption{Results of agent-based simulations of an anti-aligning Vicsek-like model. Following the nomenclature of ref.~\citenum{ihle2023}, the simulation parameters are $M=\rho \pi R^2=1$ for the reduced density (with $\rho$ the number density) and $\mathrm{Sc}=\Gamma R/v=1$ for the dimensionless coupling strength. The particle number $N$ varies $N=2^{5},\ldots,2^{12}$ which is also indicated by symbol brightness and size (smaller, darker circles corresponding to larger systems). Large: The structure factor $S(k)$ as a function of the wavenumber $k$ {in units of the inverse of the interaction length $R$}. Inset left: The apparent anomalous scaling exponent $k^{\approx 0.5}$ found from a direct fit can be explained by an exponent of two and an offset as in the 1D model, here $\Delta S=S(k)-S(0)$ with inferred $S(0)$. Squares correspond to results from half the running time to establish convergence. Inset right: The number fluctuations seen by comparison of windows of size $r$; here, we rescaled in such a way that the apparent slope is expected to be the negative of the one seen in the structure factor~\cite{torquato2018}. For $r\approx L$, the number fluctuations are inevitably suppressed. We provide a guide to the eye for the slope in both cases. } 
    \label{fig:sf2d}
\end{figure} 

We integrate the equations of motion given above using Heun's method.~\cite{ruemelin1982} We use $\Gamma=v=R=M=1$ as parameters. Unless stated otherwise, we compute the structure factor after a dimensionless time of $t=500\pi$ has passed and present results averaged over $10^4$ samples. Numerically, it is also easy to address the influence of noise. Both positional and orientational noise will eventually break long-range correlations. The case of orientational noise is the more pertinent here as we are interested in the manifestation of the alignment interactions. We can roughly estimate the critical noise level by estimating when the noise changes the interaction topology. The typical timescale between interactions is given by the mean free path and found to be $\tau_{\text{free}}\sim R/(M v)$. If the active particles move by an amount of $\ell \sim R$ transverse to the deterministic direction in that time due to diffusion, they will interact with different particles. This reasoning leads to an expectation of $D_c \approx M/\pi v R$ for the critical orientational diffusion strength. We note that this scales the same way as the threshold of the instability reported in ref.~\cite{escaff2024}; however, the data here is operating below that instability.

\begin{figure}
    \includegraphics[scale=0.5]{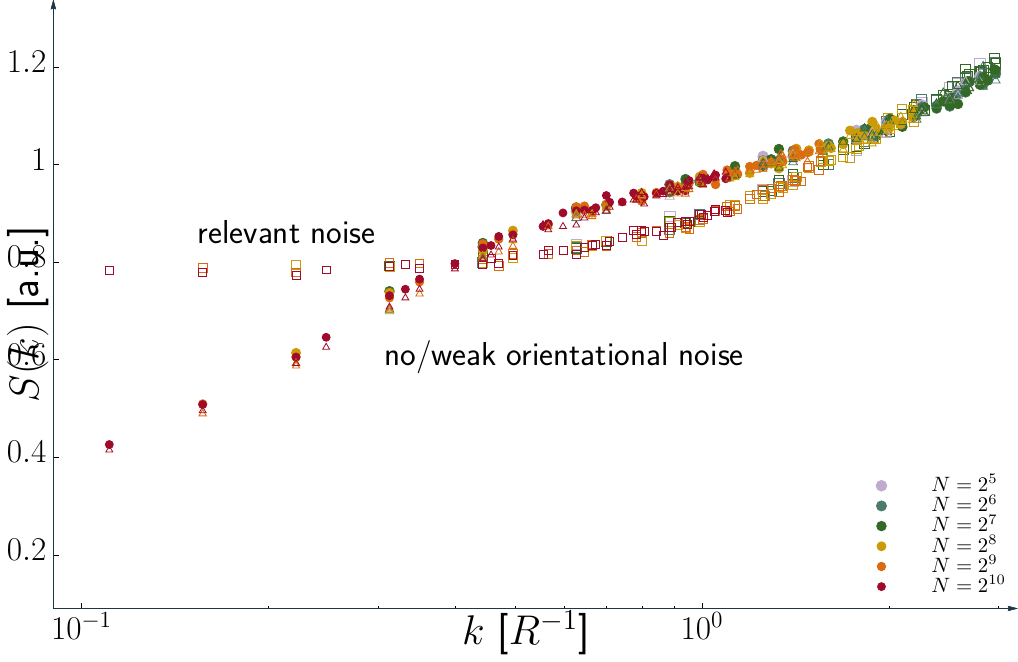}
    \caption{Demonstration of the diffusive breakdown of the long-range suppression of density-fluctuations by comparing the structure factor for a system with no orientational noise (full circles). The noise is implemented as additional white random forces $\mu \eta_i$ with $\langle \eta_i \rangle=0$ and $\langle \eta_i(t)\eta_j(t')\rangle=\delta_{ij}\delta(t-t')$. Here, we used $\mu=1$ as relevant noise (empty squares) and $\mu=0.1$ as weak noise (empty triangles) and compare them to the deterministic case (full circles).}
    \label{fig:sfnoise}
\end{figure}

\begin{figure}
    \includegraphics[scale=0.5]{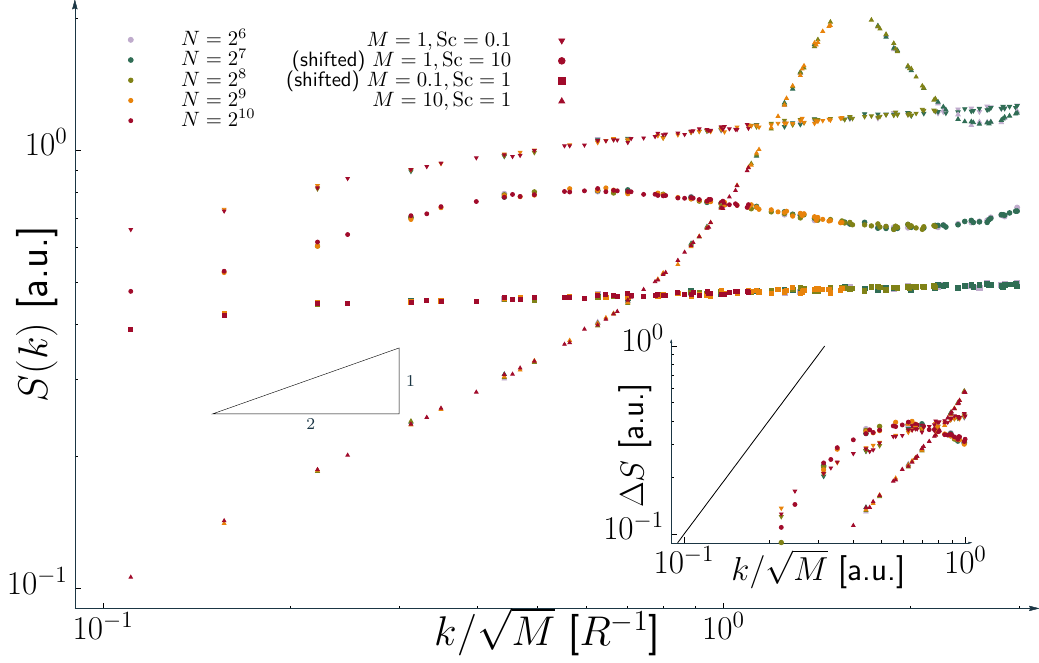} 
    \caption{Further exploration of the parameter space corroborating the conclusions drawn in the main text. Large: Additional data of the structure factor, system size is again indicated by color and parameter set by the choice of symbol. Some of the curves have been shifted vertically to enhance visual clarity. Similarly the wave number has been rescaled by system size to allow for a joint presentation. For comparison, we indicate the slope found for $M=1, \mathrm{Sc}=1$, as presented in the main text. All data show a clear reduction of density fluctuations at large distances, but the apparent power-law behaviour would be non-universal. Inset: Reduced structure factor $\Delta S=S-S_0$. $S_0$ has been inferred by fitting the model small wave-number prediction $S-S_0\propto k^2$ to the data (this was not meaningfully possible for the $M=0.1, \mathrm{Sc}=1$ data set). The solid black line corresponds to a monomial $\propto k^2$ which is provided as a guide to the eye. }
    \label{fig:sfparams}
\end{figure}

We present the numerical results from direct agent-based simulations in fig.~\ref{fig:sf2d}. We observe seemingly consistent apparent power-law behaviors in the reduced structure factor and the number fluctuations. However, one has to be wary that this can be a finite-size artifact and, indeed, a description similar to the one-dimensional model prediction of eq.~\eqref{eq:pred} does also fit the data. This is not too surprising on an algebraic level, but it is a reminder of the limitations that extrapolation of the structure factor to $k\to 0$ has. Similar ostensible hypperuniformity $S\sim k^\alpha$ with ``anomalous'' (non-integer) exponents $\alpha \approx 0.5$ has been found in systems of chiral self-propelled particles~\cite{huang2021,zhang2022,chen2024}. For the purposes of this work, we are mainly interested in a significant reduction of long-ranged density fluctuations. The question of actual hyperuniformity in this or similar systems is best addressed analytically which is currently outstanding, but the one-dimensional results should serve as a warning in this regard.  The observed reduction is robust against small noise and breaks down for large noise. In Fig.~\ref{fig:sfnoise}, we present data corroborating the intuitive notion that a sufficiently strong noise will indeed lead to a break-down of the reported effect.In Fig.~\ref{fig:sfparams}, we show data analogous to that of {Fig.~\ref{fig:sf2d}} for drastically different values of expected interaction number $M=\rho \pi R^2$ and dimensionless interaction strength $\mathrm{Sc}=\Gamma R/v$. We find, to the limitations of the system sizes studied, a consistent picture: ostensible hypperuniform behaviour with an anomalous non-integer power-law that can reasonably well be brought into the form $S(k)-S_0\propto k^2$. 

While the self-mixing dynamics in the anti-aligning model might on large scales appear similar to the effect of thermal noise, giving rise to ideas such as effective temperatures~\cite{hecht2024}, the absence of the long-ranged correlations in the regime of sufficiently strong explicit noise is a tangible difference. These findings serve as a reminder that the nature of non-equilibrium steady states in active matter systems, especially those with phase-space compression~\cite{ihle2023,boltz2023} is complex and intriguing.

\section{Discussion}

To summarize, we investigated the concept of reduced density fluctuations on long distances, possibly leading up to {or unfortunately mimicking} the formation of hyperuniform states, in active matter systems with anti-aligning orientational interactions. We rationalized the possibility from a general point of view and established the mechanism analytically in a novel one-dimensional, active lattice gas model. For this model, we determine the structure factor by solving the Langevin dynamics associated to the Poisson representation of the full probability density function. The structure factor exhibits a notable reduction of the long-range density fluctuations with a finite offset at vanishing wave-number that is at least partially rooted in the geometrical limitations of being in one-dimension.  We corroborate the general findings with direct simulations of a two-dimensional continuous system for which we find ostensible hypperuniformity. Lacking a full analytical solution of the structure factor for this more complex model, we cannot at the current time address the question of a residual offset at smaller wave-numbers that would change the leading exponent to a more conventional exponent of two. This work, however, does offer a glimpse that this scenario is a fairly relevant one, and comparison with experiments on anti-aligning self-propelled particles, e.g. along refs.~\cite{nourhani2021,das2024}, will be interesting. Aside from the question at hand, this work also establishes anti-aligning self-propelled particles as a standalone dynamical process to generate disordered point-patterns with tunable statistics. The existence of a transition from Poissonian statistics for a passive system at $v=0$ to non-Poissonian, possibly hyperuniform, statistics for active systems $v>0$ is a remarkable manifestation of the relevance of long-range correlations even in the absence of direct order.  This suppression of density fluctuations is purely a result of dynamically built-up correlations due to aligning interactions and not a result of more conventional mechanisms such as elasticity. It is this long-range effect that the one-dimensional, analytically solvable toy model is capable of capturing and that connects it to the more complex and relevant models such as the two-dimensional Vicsek-like model that we studied numerically.  Numerical inference of hyperuniformity without a principle theoretical foundation is hard and we see our work as one that challenges over-eager excitement over ostensible hypperuniformity in data in favor of a slightly less dramatic but still interesting effect that is theoretically accessible.

\bibliography{lit.bib}

\end{document}